\documentclass[twocolumn,prl,preprintnumbers,amsmath,amssymb,floatfix]{revtex4}
\usepackage[linktocpage,bookmarksopen,bookmarksnumbered]{hyperref}
\usepackage{graphicx}
\usepackage{dcolumn}

\usepackage{amsmath,graphics,epsfig,color,verbatim,ulem}

 \newcommand{\vk}{{\mathbf{k}}}

\newcommand{\Tr}{\mathrm{Tr}}

 \newcommand{\up}{\uparrow}

\newcommand{\dn}{\downarrow}
\renewcommand{\up}{\uparrow}

\begin{document}

\setlength{\pdfpageheight}{\paperheight}
\setlength{\pdfpagewidth}{\paperwidth}

\title{Fractional power-law behavior and its origin in iron-chalcogenide and ruthenate superconductors: Insights from first-principles calculations}
\author{Z. P. Yin}
 \email{yinzping@physics.rutgers.edu}
\author{K. Haule}
\author{G. Kotliar}
\affiliation{Department of Physics and Astronomy, Rutgers University, Piscataway, NJ 08854.}
\date{\today}
%%%%%%%
\begin{abstract}
We perform realistic first-principles calculations of iron
chalcogenides and ruthenate based materials to identify
experimental signatures of Hund's coupling induced correlations in these
systems. We find that FeTe and K$_x$Fe$_{2-y}$Se$_2$ display unusual orbital dependent 
fractional powerlaw behavior in
their quasiparticle self energy and optical conductivity, 
a phenomena first identified in SrRuO$_3$.
Strong incoherence in the paramagnetic state of these materials
results in electronic states hidden to angle-resolved
photoemission spectroscopy which reemerge at low temperatures.
We identify the effective low energy Hamiltonian describing these
systems and show that these anomalies are not controlled by the
proximity to a quantum critical point but result from
coexistence of fast quantum mechanical orbital fluctuations and
slow spin fluctuations.
\end{abstract}
\maketitle

\section{Introduction}

The study of the Hund's coupling effects in solids has a long
history. Van der Marel and Sawatsky\cite{Marel} pointed out that
unlike the Hubbard $U$ which is strongly screened, the atomic Hund's
$J_H$, persists essentially unrenormalized in the solids and increases
the splittings between the lower and the upper Hubbard bands for a
half filled shell, while decreases it away from half filling.  The
Hund's term was also shown to have important consequences on the low
energy physics of quasiparticles, when a transition metal impurity is
screened in metallic host. The Hund's coupling was shown to
dramatically reduce the value of the Kondo
temperature\cite{Schrieffer,Okada}. Recent interest in this problem
arose from dynamical mean field theory\cite{DMFT-RMP1996} (DMFT)
investigations of the recently discovered iron pnictide
superconductors. It was proposed that in these materials strong
correlations arise from the Hund's rule coupling $J_H$\cite{Haule-njp}
rather than from the Hubbard $U$, resulting in large mass
enhancements. These calculations\cite{Haule-njp} showed that for a
reasonable value of the Hubbard $U$, the mass enhancement due to
interactions is very small when $J_H$=0, whereas it is exponentially
enhanced by the Hund's rule coupling.  Optical spectroscopy studies
have shown that in both iron pnictides and chalcogenides the optical
masses are many times larger than the band
masses\cite{optics-LaFePO,optics-LaFeAsO,Degiorgi,Basov-optics}.  The
trend in mass enhancements is well accounted for by DMFT combined with
density functional theory (DFT+DMFT) calculations\cite{Yin-nm}. Since
the strength of correlations in these solids is almost entirely due to
the Hund's coupling, these materials are dubbed Hund's
metals\cite{Yin-nm}. The role of Hund's coupling in iron pnictides and
chalcogenides has been addressed from different perspectives in the
literature.~\cite{Haule1,Yin-np,Mazin,Yin-nm,XDai,held,ZWang,WYin,Liebsch,QSi,LCraco,Schichling,
  JFerber, Misawa, NLWang, PPhillips, WYin3} Powerlaw behavior in
quasiparticle self energy of model Hamiltonians with Hund's coupling
was discovered in Ref.~\cite{Werner2} and related to observations in
ruthenates\cite{optics-SRO4,optics-SRO3}.  Many anomalous properties
of ruthenates\cite{Mravlje} and other $4d$ compounds were shown to be
governed by Hund's physics.~\cite{Medici2}

While at low energies and low temperatures Hund's metals are
describable by Fermi liquid theory, the physical properties in
their incoherent regime are anomalous and surprising. In the
iron pnictides and chalcogenides there is a strong tendency
towards orbital differentiation~\cite{Yin-nm}, and the large mass
enhancement can occur even though no clear Hubbard band exist in
the one particle spectra of these Hund's metals~\cite{Kutepov}.

In this article, we use first principles methods and model
Hamiltonians to search for experimental signatures of Hund's
physics in iron chalcogenides and ruthenates which are the subject of 
current intensive experimental studies. 
We show how the incoherence in iron chalcogenides above the
N\'{e}el temperature can blur portions of the Fermi surface
rendering them dark to photoemission spectroscopy. 
We show that fractional powerlaw behavior in optical
conductivity that received significant attention in the
ruthenates also takes place in the FeTe system, deepening the
analogies between these systems. We compare the powerlaw exponents in
optical conductivity extracted from first principles DFT+DMFT
calculations with experiments in a broad class of materials, and
elucidate the control parameters that govern this behavior. The fractional
powerlaw behavior is characteristic of an intermediate regime where the
orbital degrees of freedom are quenched but the spin degrees of
freedom are not. This physics is most pronounced at the special
valence of one unit of charge away from the half-filling.

\section{Results}

\subsection{DFT+DMFT results}

\begin{figure}[!ht]
\centering{
\includegraphics[width=0.99\linewidth]{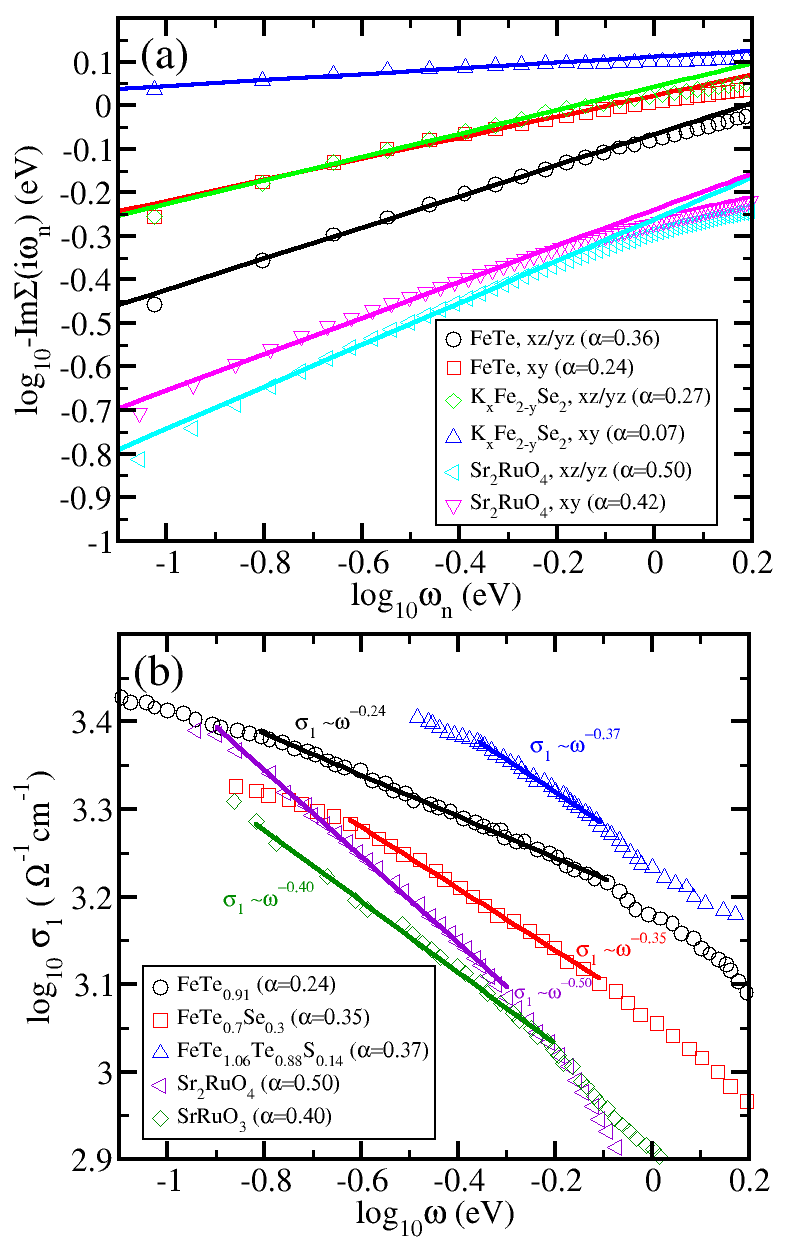}
 }
\caption{
(color online) Fractional powerlaw in (a) theoretical self-energy and (b) experimental optical conductivity in iron chalcogenides and ruthenates.
Experimental data are taken from ref. \cite{optics-FeTe1} for FeTe$_{0.91}$ and FeTe$_{0.7}$Se$_{0.3}$,
ref.~\cite{optics-FeTe2} for Fe$_{1.06}$Te$_{0.88}$S$_{0.14}$, ref. \cite{optics-SRO4} for Sr$_2$RuO$_4$,
and ref. \cite{optics-SRO3} for SrRuO$_3$.
}
\label{powerlaw}
\end{figure}

We first show results of our realistic DFT+DMFT
calculation~\cite{DMFT-RMP2006, Haule-DMFT} for three compounds
currently under extensive investigations: ruthenates Sr$_2$RuO$_4$,
and iron chalcogenides FeTe and K$_{x}$Fe$_{2-y}$Se$_2$. In all the
DFT+DMFT calculations, we use the same Coulomb repulsion $U$=5.0 eV
and Hund's $J_H$=0.80 eV, as determined from \textit{ab initio} in
Ref.~\onlinecite{Kutepov}, and used in our previous work~\cite{Yin-np,
Yin-nm}. 
Notice that we include all the electronic states in 
a large energy window (-10 eV to 10 eV) as opposed to 3- or 5-band model 
calculations, thus the Hubburd $U$ appropriate for our calculations is much 
less screened compared to those studies.  
The electronic charge is computed self-consistently on
DFT+DMFT electronic density. The quantum impurity problem is solved by
the continuous time quantum Monte Carlo (CTQMC)
method~\cite{Haule-QMC,Werner}, using Slater form of the Coulomb
repulsion in its fully rotational invariant form. We use the
experimentally determined lattice structures, including the internal
positions of the atoms for Sr$_2$RuO$_4$~\cite{a-SRO},
FeTe~\cite{a-FeTe}, K$_{x}$Fe$_{2-y}$Se$_2$~\cite{a-KFS}, and
FeSe~\cite{a-FeSe}.

We show in Fig.\ref{powerlaw}(a) the imaginary part of the self-energy
of the $t2g$ orbitals in Sr$_2$RuO$_4$, FeTe and
K$_{x}$Fe$_{2-y}$Se$_2$ on the imaginary axis, plotted in log$_{10}$
scale. In the intermediate energy range from a low energy cutoff
$\sim$0.1 eV, below which the materials gain coherence, to roughly
Hund's $J_H$$\sim$0.8 eV, the imaginary part of the self-energy
clearly shows a fractional powerlaw behavior, i.e.,
$Im\Sigma(i\omega_n)\propto-\omega_n^\alpha$. For the normal Fermi
liquid, this exponent is unity, and at finite temperature correlated
materials have additional constant scattering rate. The fractional
powerlaws are however very uncommon.

From the quantum chemistry perspective, both iron chalchogenides
and ruthenates share a common theme: they contain correlated
electrons with the $d$ valence of one unit charge away from
half-filling. In iron pnictides/chalcogenides, the Fe-ion is
surrounded by a tetrahedron of pnictogen/chalcogen, and the
resulting crystal field splittings are very small compared to
Fe-pnictogen hybridization~\cite{Haule-njp} hence all 5 Fe 3$d$
orbitals are active. Their average occupancy is close to $d^6$,
one unit of charge away from the half-filled $d^5$. For the
ruthenates, the coordination of the Ru is octahedral, and the
oxygen ligands induce a large $t2g$-$eg$ splitting, with only the
$t2g$ orbitals active, containing approximately $4$ electrons in
three $t2g$ orbitals, one electron more than the half filled
shell.

The values of the apparent powerlaw exponents differ from material to
material and deviate even for different orbitals of the same
material, which is connected to the orbital occupancy. As shown in
Fig.~\ref{powerlaw}(a), the $xz/yz$ orbitals of Sr$_2$RuO$_4$
show an exponent of 0.5, while the more correlated $xy$ orbital,
which is closer to half-filling, show a smaller exponent of 0.42
. In iron pnictides and chalcogenides, the average occupancy per
orbital is even closer to half-filling (only $1/5$ away, as
opposed to $1/3$ in ruthenates). As we will show below by a model
study, one expects stronger electronic correlations in this case
and a smaller powerlaw exponent. Indeed, the $xz/yz$ orbitals in
FeTe show exponent of $\approx 0.36$ whereas for the more strongly
correlated $xy$ orbital, the exponent is only $\approx 0.24$.
Iron vacancies in the
K$_{x}$Fe$_{2-y}$Se$_2$ makes the compound even more
correlated than FeTe, and the powerlaw exponent is further reduced to
0.27 for the $xz/yz$ orbital, and only 0.07 for the $xy$ orbital.

The powerlaw behavior of the self-energy manifests itself in optical
conductivity studies. In a simplified treatment, the optical
conductivity can be approximated by $\sigma(\omega)\propto
Re(1/(\omega+i\Sigma''(\omega)+\Sigma'(\omega)-\Sigma'(\omega=0)))$~\cite{Basov-review}.
In Fig.~\ref{powerlaw}(b), we present experimental data on
FeTe$_{0.91}$\cite{optics-FeTe1},
FeTe$_{0.7}$Se$_{0.3}$\cite{optics-FeTe1},
Fe$_{1.06}$Te$_{0.88}$S$_{0.14}$\cite{optics-FeTe2},
Sr$_2$RuO$_4$\cite{optics-SRO4}, and SrRuO$_3$\cite{optics-SRO3}.
As can be seen in Fig.~\ref{powerlaw}(b), the optical
conductivity in these materials can be roughly approximated by
$\sigma_1(\omega)\sim B\omega^{-\alpha}$ in about the same energy
range as in the theoretical self-energy. The experimental
exponents obtained from optical conductivity are very similar to
the theoretical exponents for the self-energy, as expected from
the simplified relation between optical conductivity and
self-energy.

\begin{figure}[t]
\centering{
\includegraphics[width=0.99\linewidth]{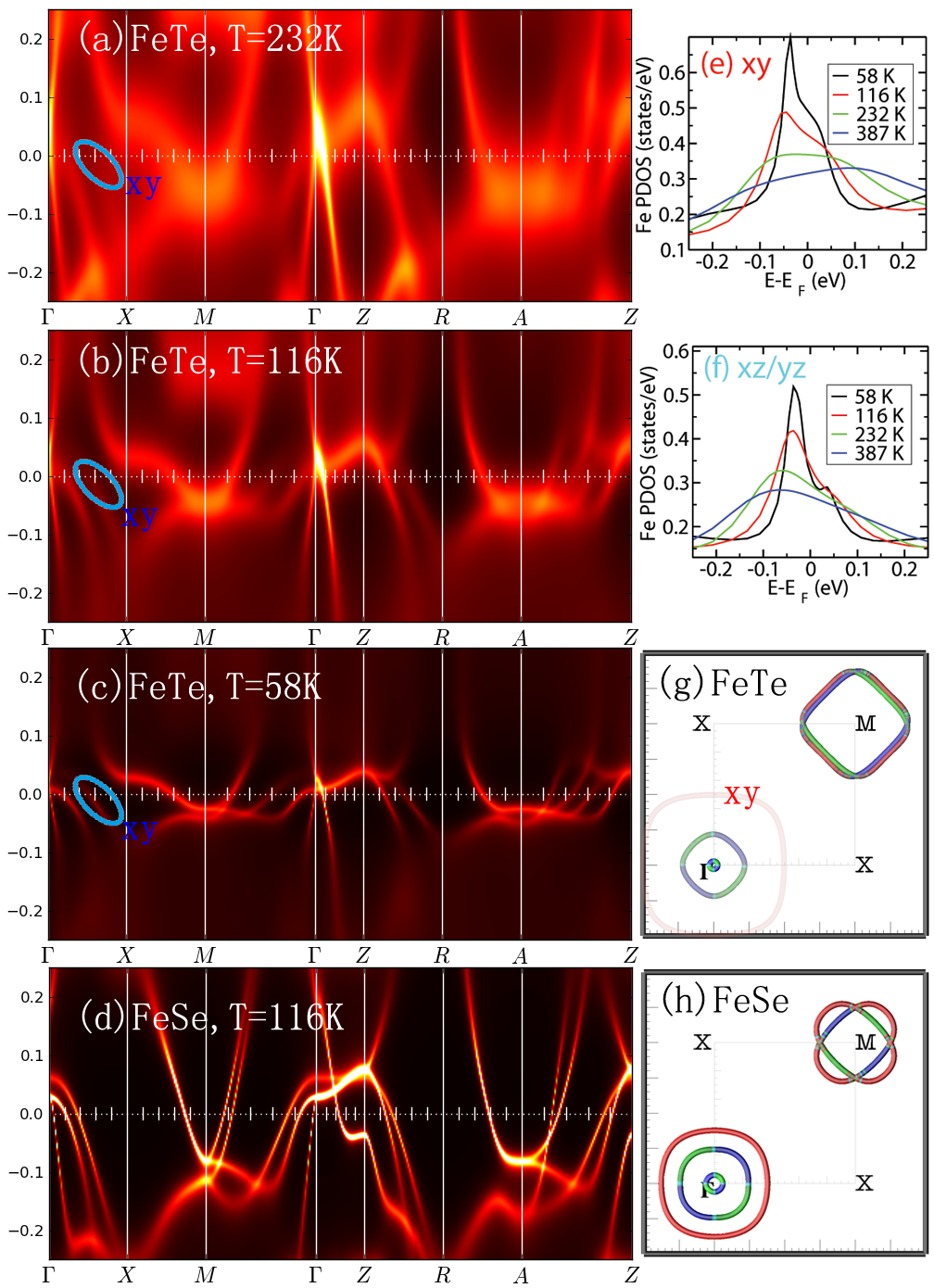}
 }
\caption{
(color online) Incoherence-coherence crossover in FeTe.
$A(\vk,\omega)$ of along the path
$\Gamma \rightarrow X \rightarrow M\rightarrow \Gamma \rightarrow Z \rightarrow R \rightarrow A\rightarrow Z$
for FeTe at (a) 232 K, (b) 116 K, and (c) 58 K and for (d) FeSe at 116 K in the PM states.
(e) and (f) $A(\omega)$ for Fe $3d$ $xy$ and $xz/yz$ orbital at 387 K, 232 K, 116 K and 58 K in PM FeTe.
(g) and (h) color-coded Fermi surface in the $\Gamma$ plane for PM FeTe and FeSe, respectively.
Red, green and blue color correspond to $xy$, $xz$ and $yz$ orbital character, respectively.
Due to the incoherent nature of the $xy$ orbital above T$_N$, the outer hole pocket around $\Gamma$ is not easy to be detected in ARPES experiment.
}
\label{incoherence-crossover}
\end{figure}

Hunds metals have a very low temperature scale, called the coherence
temperature, below which a Fermi liquid-like coherence regime is reached. This
phenomena has been discussed in other contexts such as heavy
fermions\cite{Fisk, Reinert} and transition metal
oxides\cite{Valla} and can be fruitfully probed by
photoemission spectroscopy. At finite temperatures, some
electronic states can be very incoherent and coherence in
different electronic states is usually not reached
simultaneously. 
Due to the strong orbital differentiation discussed in Ref.~\cite{Yin-nm},
the $t_{2g}$ orbitals have lower coherence temperature than the 
$e_g$ orbitals in the iron-based superconductors. 
Within the $t_{2g}$ shell, the $xy$ orbital has the lowest coherence 
temperature. In Fig.~\ref{incoherence-crossover} we show the
gradual evolution of the Fe $t_{2g}$ orbitals 
from very incoherent state at high temperature
to partially coherent state at lower temperature 
in paramagnetic (PM) state of FeTe. We display the momentum/orbital resolved
density of electronic states at temperatures of 387~K, 232~K,
116~K and 58~K. For comparison, we also show momentum resolved
density of state of PM FeSe at 116 K, where all electronic states
are quite coherent. The buildup of coherence in orbitally
resolved spectra of FeTe is seen as a gradual buildup of the
quasiparticle peak from a broad hump at elevated temperature to
sharper peak at lower temperature in
Figs.~\ref{incoherence-crossover}(e-f). In momentum space, the
coherence is achieved more unevenly. While some bands can be
identified at 116K, and become pretty sharp at 58K, other bands
are barely noticable even at 58K. In particular, the band of
primarily $xy$ character circled by blue ellipse, has enormous
scattering rate at 58K and should be hard to be detected by
ARPES. The missing Fermi surface is drawn in
Fig.~\ref{incoherence-crossover}g as large red pocket centered at
$\Gamma$ point, which is very incoherent above T$_N$, hence
missing in the photoemission of the paramagnetic FeTe, in strong
contrast to paramagnetic FeSe (Fig.~\ref{incoherence-crossover}h,
see also Ref.~\cite{FeSe-Biermann}).
Our calculation shows that K-intercalated FeSe
(K$_{x}$Fe$_{2-y}$Se$_2$) is even more correlated than FeTe, has
smaller powerlaw exponents, and lower coherence temperature than FeTe.
This is in agreement with recent angle-resolved photoemission
spectroscopy (ARPES) experiments on $A$$_{x}$Fe$_{2-y}$Se$_2$
compounds ($A$=K, Rb, Cs) where orbital dependent
incoherence-coherence crossover was observed by Yi, Shen and
collaborators\cite{Yi}.

\subsection{Low energy Hamiltonian}

To gain some understanding of the Hund's physics in these systems, we
derive below a low energy Hamiltonian of the three band Hubbard model,
the simplest model which shows powerlaw behavior of the self-energy. The
starting Hubbard Hamiltonian is $H=H_t+H_U$, with the hopping term
$H_t=\sum_{ij\sigma, a,b} t^{ij}_{a b} f^\dagger_{i a \sigma} f_{j b \sigma}$
and Coulomb term
$H_U=\frac{1}{2}\sum_{i\sigma,a b c d} U[a,b,c,d] f^\dagger_{i a \sigma} f^\dagger_{i b \sigma'}f_{i c \sigma'}f_{i d \sigma}$.
Here $a,b,c,d$ ($i,j$) are orbital (site) indices, and $\sigma$ stands
for the spin.
The hopping
term is taken to be locally SU(6) symmetric (no crystal fields), while the Coulomb
interaction is set to
$U[a,b,c,d]=U\delta_{ad}\delta_{bc}+J\delta_{ac}\delta_{bd}$, which
reduces the symmetry to SU(3)$\times$SU(2).
Within DMFT, this model maps to
a SU(3)$\times$SU(2) impurity Hamiltonian.
To understand why the Hund's rule coupling has such a dramatic effect
on the physical properties, we
first perform Schrieffer-Wolff transformation (for derivation see appendix) 
to obtain Kondo-like
Hamiltonian, of the form $H^{Kondo}_{eff}=H_0+H_1+H_2+H_3$, with
the potential scattering term
$H_0 = J_{p} \sum_{a \sigma}{\psi_{a \sigma}^\dagger(0)} \psi_{a \sigma}(0)$,
the spin-spin Kondo part
$H_1 = J_{1} \sum_\alpha S^\alpha \sum_{a \sigma \sigma'}{\psi^\dagger_{a \sigma}}(0) {\sigma^\alpha}_{\sigma \sigma'}\psi_{a \sigma'}(0)$,
the orbital-Kondo part,
$H_2 = J_{2} \sum_\alpha T^\alpha \sum_{a b \sigma}{\psi^\dagger}_{a \sigma}(0) {\lambda^\alpha}_{a b}\psi_{b \sigma}(0)$,
and the coupled spin-orbital part
$H_3 = J_{3} \sum_{\alpha\beta} T^\alpha S^\beta \sum_{a b \sigma} {\psi^\dagger}_{a \sigma}(0) {\lambda^\alpha}_{a b} {\sigma^\beta}_{\sigma \sigma'}\psi_{b \sigma'}(0).$
Here $S^\alpha=\sum_{a\sigma\sigma'} f^\dagger_{a\sigma}\frac{1}{2}\sigma^\alpha_{\sigma \sigma'} f_{a\sigma'}$ and
$T^\beta=\sum_{a b\sigma} f^\dagger_{a\sigma} \lambda^\beta_{a b} f_{b\sigma'}$
are spin, and SU(3) orbital operators
acting on the impurity site, $\psi(0)$ are field operators of the
conduction electrons coupled to the impurity,
while $\sigma^\alpha_{\sigma\sigma'}$ and $\lambda_{a b}^\alpha$ are Pauli
matrices and the Gell-Mann $3\times 3$ matrices of the
 SU(3) group, respectively.

Notice that in our picture the same electrons carry both orbital and
spin degrees of freedom, in contrast to the point of view of
Ref.~\onlinecite{Coleman}, which emphasizes the spin and
orbital degrees of freedom being carried by different type of
electrons, i.e., $t2g$ the spin, and $eg$ the orbital.

While the form of the low energy impurity model is dictated by
symmetry considerations, the exchange couplings $J_1,J_2,J_3$ depend
crucially on the impurity valence and Hund's coupling $J_H$. 
For the half-filled shell and large $J_H$, only
the spin-spin term $J_1$ survives, and a well known reduction of the
$J_1$ Kondo coupling for a factor of $(2l+1)$ was derived in
Refs.~\onlinecite{Schrieffer,Okada} compared to a corresponding one
band model.
Consequently, a huge reduction of the
Kondo temperature for a factor of $(2l+1)^2$ in the exponent was
derived in Ref.~\onlinecite{Okada}. This regime is relevant for
half-filled $d^5$ shell realized in the
Hund's insulators LaMnPO~\cite{Simonson}.

For the above Hund's metals, 
the relevant valence of transition
metal ion is one unit of charge away from half-filling. 
When $J_H$ is negligible, the Hamiltonian is SU(6) symmetric, and
all three Kondo couplings $J_1,J_2,J_3$ are positive
(antiferromagnetic). For the valence $n_{imp}=2$ (or $n_{imp}=4$),
their numerical values are $J_1=J_0/3$, $J_2=J_0/4$, and $J_3=J_0/2$,
where $J_0=V^2/(2 U+\varepsilon_f)$ (or $J_0=V^2/(3 U+\varepsilon_f)$) 
is a positive number, which depends on the corresponding Anderson
impurity model parameters, i.e., hybridization $V$ and impurity level
$\varepsilon_f$. The ground state is a Fermi liquid, because
antiferromagnetic couplings between conduction electrons and impurity
degrees of freedom ensure complete quenching of both the orbital and
spin moments. On the other hand, when $J_H$ is large, the
spin-spin Kondo coupling $J_1$ changes sign to ferromagnetic, while
the orbit $J_2$ and spin-orbit $J_3$ remain positive. In the three
band SU(2)$\times$SU(3) model and large $J_H$, their numerical
values are $J_1=-J_0/9$, $J_2=J_0/3$, and $J_3=J_0/3$, where
$J_0=V^2/(2 U-2 J_H+\varepsilon_f)>0$ (or $J_0=V^2/(3
U+J_H+\varepsilon_f)>0$) for valence $n_{imp}=2$ (or $n_{imp}=4$).
This change of sign is due to the orbital blocking
mechanism~\cite{Yin-nm}, that allows only those virtual charge
excitations which go through orbital singlet intermediate state (see appendix).
We note that for valence $d^6$ in iron pnictides/chalcogenides,
the correct low energy Hamiltonian has three terms, but not just
spin-spin term, as proposed earlier~\cite{Nevidomskyy}. It is however
the spin-spin $J_1$ term that changes sign in the limit of large
$J_H$, and impedes quenching of the spin degrees of freedom
(termed spin freezing in Ref.~\onlinecite{Werner2}). This
substantially reduces coherence temperature, however, the $J_3$ term,
which couples spin and orbital, is positive and gives rise to the
Fermi liquid state at very low temperature.

\begin{figure}[t]
\centering{
\includegraphics[width=0.99\linewidth]{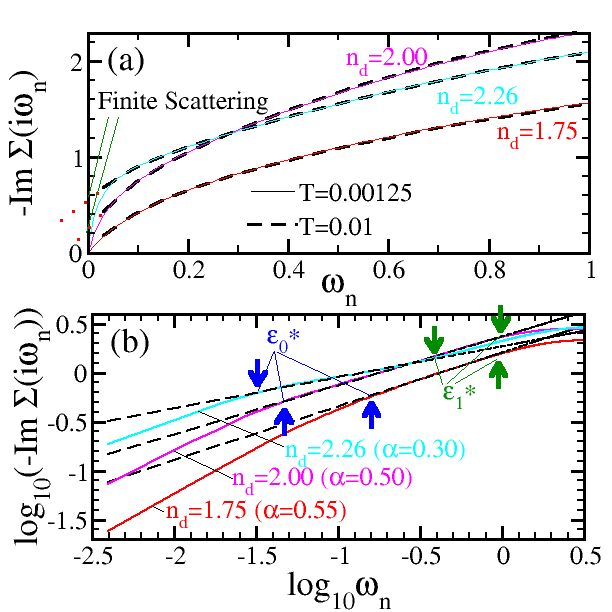}
}
\caption{
(color online) Quasiparticle self energies for a three band model with $J_H=2.0$ at different filling $n_d$=1.75, 2.00, and 2.26.
(a) The self energies at two temperatures T=0.01 and 0.00125 show the incoherence-coherence crossover with decreasing temperature.
(b) The self energies plotted in log$_{10}$ scale display fractional powerlaw behavior in the intermediate frequency range 
from $\varepsilon_0^*$ to $\varepsilon_1^*$ as indicated by arrows.
}
\label{model-self-energy}
\end{figure}

\subsection{Model Hamiltonian calculations}

To demonstrate the above picture, we numerically solve a simplified three band
model with the nearest neighbor diagonal hopping
$t_{\alpha\alpha}=0.4$, and the next nearest neighbor hopping
$t'_{\alpha\alpha}=0.4$ and $t'_{\alpha\ne\beta}=0.2$, which
gives a total bandwidth of the tight banding model $W \approx
3.5$. We take $U=6$ and large Hund's coupling $J_H$=2 and $J_H=1$ for
powerlaw to extend over a larger frequency range. 

\begin{figure}[!ht]
\centering{
\includegraphics[width=0.99\linewidth]{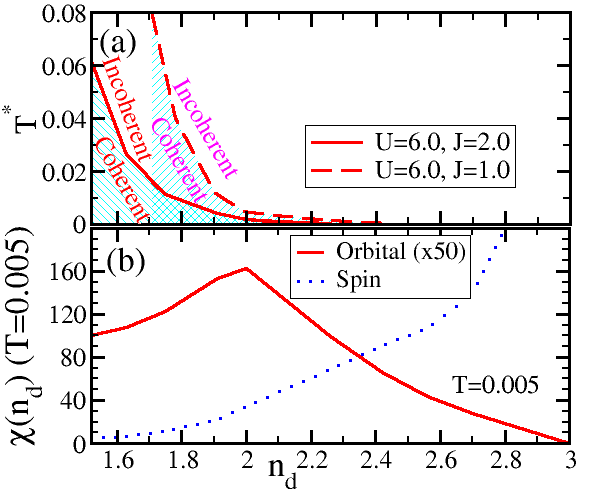}
}
\caption{
(color online) (a) The coherent temperature;
and (b) the spin and orbital susceptibility,
as a functional of $n_d$ for a three band model with $J_H=2.0$ (solid lines) and $J_H=1.0$ (dash lines).
}
\label{model}
\end{figure}

In Fig.~\ref{model-self-energy}a, we show the imaginary part 
of the quasiparticle self energies for $J_H$=2.
At the intermediate temperature T=0.01, the self energies of 
$n_d$=2.00 and 2.26 display finite values at zero frequency by extrapolation, 
which suggests incoherent properties at this temperature. 
However, at a lower temperature T=0.00125, the corresponding self energies 
clearly display Fermi liquid behavior at low frequencies. 
Therefore there is an incoherence-coherence 
crossover with decreasing temperature. 
We determine the coherence temperature as 
the temperature at which the renormalized
scattering rate is equal to the temperature, i.e.,
$-zIm\Sigma(i\omega=i0^+,T^*)=k_B T^*$, where $1/z=1-\partial
Im\Sigma(i\omega=i0^+, T^*)/\partial\omega$ and $k_B$ is the
Boltzmann constant. 
We show the coherent temperature in Fig.~\ref{model}a 
as a function of electron occupation $n_d$.
We reached \textit{8-times} lower temperature than previous studies~\cite{Werner2} 
to access the Fermi liquid state at filling far
beyond $n_d=2$, and map out the coherence incoherence
crossover temperature $T^*$.

A good powerlaw fit to the self-energy, as shown in Fig.~\ref{model-self-energy}b, 
is obtained only in a
limited range of frequency between the low energy cutoff
proportional to the Fermi liquid scale ($\varepsilon_0^*$ in
Fig.~\ref{model-self-energy}b), and the high energy cutoff ($\varepsilon_1^*$),
which is always smaller than the Hund's coupling. 
The range of frequencies at which the powerlaw is valid
($\varepsilon_0^*<\omega_n<\varepsilon_1^*$, as indicated by arrows in Fig.~\ref{model-self-energy}b) 
is largest at valence $n_d=2$, 
where the exponent is close to $1/2$, as
previously reported in Ref.~\cite{Werner2}. 
Much lower temperature
reached in this work show that exponent $\alpha$ decreases monotonically
with increasing $n_d$ (see Fig.~\ref{model-self-energy}b and Figs.~\ref{self1} and \ref{self2} in appendix), 
in contrast to Ref.~\cite{Werner2}, 
hence stronger correlations
approaching the Mott state at $n_d=3$ lead to smaller exponent at
low temperature, in qualitative agreement with our realistic
calculations for Fe and Ru compounds. Most importantly, there is
no signature of quantum phase transition to non-Fermi liquid
spin-frozen state around valence $n_d=2$, where the powerlaw
exponents are found, and the crossover temperature scale does not
follow the powerlaw behavior $T^*\sim(x-x_c)^{z \nu}$ expected in
a quantum critical scenario. Hence we can \textit{exclude} the possibility
that the exponents are due to the proximity to the quantum phase
transition, as proposed in Ref.~\cite{Werner2, Biermann}. 

\section{Discussions}

The powerlaw
exponents are found in the temperature and frequency regime where
the spin degrees of freedom are very slow and unquenched (spin
susceptibility has Curie-Weiss form and large static value) while
the orbital degrees of freedom are very fast and quenched (orbital
susceptibility is Pauli-like, but enhanced).(see Fig.~\ref{model}b and Fig.~\ref{susc} in appendix)
This is a novel regime in which two degrees of freedom behave in
different ways , one fluctuates very fast (positive $J_2$), the
other one fluctuates very slowly (negative $J_1$), and both
coupled by a third antiferromagnetic coupling $J_3$. This
situation is similar to the intermediate phase of the extended
Hubbard model study of Ref.~\cite{QSi2}, where similar
continuously varying exponents were shown to exists in the metallic
non-Fermi liquid phase in which there was a quenched spin-degree
of freedom and an unquenched charge degree of freedom. Notice
that at $n_d=3$ only the large spin state is possible, hence the
orbital degrees of freedom are gapped, and exponents disappear,
while the effect is maximal one unit of charge away from
half-filling, i.e., $n_d=2$.

\section{Conclusions}
In conclusions, we have shown in this paper that the Hund's rule coupling has a strong impact on the
electronic states in valence of one unit of charge away from
half-filling. The strongly correlated state in such materials can have
very low coherence temperature, and the self-energy and optical
conductivity show fractional powerlaw at intermediate energy. 
We have derived the effective low energy Hamiltonian describing these
systems and identified a negative Kondo coupling in the
spin-spin part of the corresponding low energy Kondo model.
We have shown that these anomalies are not controlled by the
proximity to a quantum critical point but result from
coexistence of fast quantum mechanical orbital fluctuations and
slow spin fluctuations.
This is relevant for ruthenates and iron chalcogenides, as well as many other
materials with similar valence and sizable Hund's coupling.

\section{Acknowledgements}
We thank Dimitri Basov, Antoine Georges, Jernej Mravlje, Philipp Werner and Andrew Millis for fruitful discussions.
ZPY and GK were supported by NSF DMR-0906943, KH was supported
by NSF DMR-0746395.

\section{APPENDIX}
\subsection{Schrieffer-Wolff transformation}

We start our discussion with the three band Hubbard model $H=H_t+H_U$,
with the hopping term
\begin{equation}
  H_t=\sum_{ij\sigma, a,b} t^{ij}_{a b} f^\dagger_{i a \sigma} f_{j b \sigma}
\end{equation}
and the Coulomb repulsion term
\begin{equation}
H_U=\frac{1}{2}\sum_{i\sigma,a b c d} U[a,b,c,d] f^\dagger_{i a \sigma}
f^\dagger_{i b \sigma'}f_{i c \sigma'}f_{i d \sigma}.
\end{equation}
Here index $a$ runs over the three orbitals, $i,j$ over lattice sites,
and $\sigma$ over spin.
The hopping term is taken to be locally SU(6) symmetric, while the
Coulomb interaction is set to
$U[a,b,c,d]=U\delta_{ad}\delta_{bc}+J\delta_{ac}\delta_{bd}$, which
reduces the symmetry to SU(3)$\times$SU(2).

Within the Dynamical Mean Field Theory, this model maps to the
SU(3)$\times$SU(2) impurity Hamiltonian of the form
$H_{imp}=H_{bath}+ H_{hyb} + H_{local}$
\begin{eqnarray}
H_{bath} = \sum_{\vk a \sigma} \varepsilon_{\vk a}
\psi_{\vk a\sigma}^\dagger \psi_{\vk a\sigma}
\end{eqnarray}
\begin{eqnarray}
H_{hyb}=\sum_{\vk a \sigma} V_{\vk a} \psi^\dagger_{\vk a\sigma}
f_{a\sigma} + h.c.
\end{eqnarray}
\begin{eqnarray}
H_{local}=\sum_{a\sigma} \varepsilon_f f^\dagger_{a\sigma} f_{a\sigma} + H_{U}
\end{eqnarray}
To gain further insights into the low energy degrees of freedom of
this Hamiltonian, we
perform the Schrieffer-Wolff transformation, which takes the form
\begin{equation}
H_{eff} = P_n H_{hyb} \frac{P_{n\pm 1}}{\Delta E} H_{hyb} P_n,
\label{wolf}
\end{equation}
where $P_n$ is the projector to the impurity ground state multiplet, and
$\Delta E$ is the energy cost for the charge excitation from the
ground state multiplet to the $n\pm 1$ lowest energy multiplet states, and is always negative.

Here we will limit our discussion to the case of a ground state valence
$n_f=2$ and virtual charge excitations into valence $n_f=3$.
A direct way of evaluating this effective Hamiltonian is to perform
exact diagonalization of $H_{local}$, then expressing matrix elements
of $f_{a\sigma}$ operators in terms of atomic eigenstates
$(F^\dagger_{a\sigma})_{m_1 m_2}=\langle
m_1|f^\dagger_{a\sigma}|m_2\rangle$, and finally evaluating
all terms which appear in the sum
\begin{eqnarray}
H_{eff}=\sum_{\vk\vk' a b \sigma\sigma'} \frac{V_{\vk a}V_{\vk'b}^*}{\Delta E}
\psi^\dagger_{\vk a\sigma}
\psi_{\vk' b\sigma'}\times\nonumber\\
\sum_{m_1 m_2 m_3}(F_{a\sigma})_{m_1 m_2}(F^\dagger_{b\sigma'})_{m_2 m_3}
 |m_1\rangle\langle m_3|.
\end{eqnarray}
Here $m_2$ runs over the ground state
multiplet at valence $n+1$, while $m_1$ and $m_3$ run over ground
state multiplet at valence $n$. Finally we need to express
the impurity degrees of freedom in terms of the impurity operators such as
is the total spin $S$ and the orbital isospin operator $T$.

This tedious derivation can be circumvented by a trick. We
Fourier transform the bath operators
$V^2 \psi^\dagger_{a\sigma}(0) =\sum_\vk V_{\vk a}\psi^\dagger_{\vk
 a\sigma}$ and introduce the combined spin-orbit index $i\equiv(a\sigma)$.
We can then rewrite the effective Hamiltonian for charge
excitations from valence $n$ to valence $n+1$ as
\begin{eqnarray}
H_{eff}=\sum_{i j k l} \frac{V^2}{\Delta E}
\psi^\dagger_{i}(0)
\psi_{j}(0)
P_n f_{k} P_{n+1} f^\dagger_{l} P_n \;\delta(i,k)\delta(j,l)
\end{eqnarray}
We next find a complete orthonormal basis in the space of spin and orbit degrees
of freedom ($\Tr(I^\alpha I^{\beta \dagger})=\delta_{\alpha\beta}$), in which the completeness relation takes the form
\begin{eqnarray}
\delta(i,k)\delta(j,l) = \sum_\alpha (I^{\alpha *})_{ij} (I^\alpha)_{k l}.
\end{eqnarray}
Here matrices $I^\alpha$ form a complete basis for the SU(3)$\times$SU(2)
group. For the SU(2) and SU(3) group we use Pauli $2\times 2$ matrices $\sigma^\alpha$, and
Gell-Mann $3\times 3$ matrices $\lambda^\alpha$. In terms of these, the complete
basis $I^\alpha$ is
\begin{equation}
I^\alpha = \left\{
\begin{array}{c}
\frac{1}{\sqrt{3}} 1 \otimes 1 \frac{1}{\sqrt{2}}\\
\frac{1}{\sqrt{3}} 1 \otimes \sigma \frac{1}{\sqrt{2}}\\
\frac{1}{\sqrt{2}} \lambda \otimes 1 \frac{1}{\sqrt{2}}\\
\frac{1}{\sqrt{2}} \lambda \otimes \sigma \frac{1}{\sqrt{2}}\\
\end{array}
  \right.
\label{Ias}
\end{equation}
The normalization factors come from the fact that
$\textrm{Tr}(\sigma^\alpha\sigma^\beta)=2\delta(\alpha,\beta)$ and
$\textrm{Tr}(\lambda^\alpha\lambda^\beta)=2\delta(\alpha,\beta)$.
We can then simplify the low energy Hamiltonian by
\begin{eqnarray}
H_{eff}=\sum_{i j k l,\alpha} \frac{V^2}{\Delta E}
(I^\alpha)_{k l}\; P_n f_{k} P_{n+1} f^\dagger_{l}P_n \;
\psi^\dagger_{i}(0)(I^{\alpha *})_{ij} \psi_{j}(0)
\end{eqnarray}
Next we realize that even in the presence of an arbitrary projector,
the local operators keep the same form of the expansion in terms of
the electron field operator
\begin{eqnarray}
\sum_{k l}(1\otimes \sigma^\alpha)_{k l}\; P_n f_{k} P_{n+1} f^\dagger_{l} P_n&\propto& -S^\alpha\\
\sum_{k l}(\lambda^\alpha \otimes 1)_{k l}\; P_n f_{k} P_{n+1} f^\dagger_{l} P_n &\propto& -T^\alpha\\
\sum_{k l}(\lambda^\alpha \otimes \sigma^\beta)_{k l}\; P_n f_{k} P_{n+1} f^\dagger_{l} P_n &\propto&  -T^\alpha\otimes S^\beta,
\end{eqnarray}
but the proportionality constants need to be determined by an explicit
calculation. Notice that just like in the Wigner-Eckart theorem, we
only need to consider one matrix element to determine proportionality
constant, which greatly simplifies this derivation.

Now we can recognize that the first term in Eq.~(\ref{Ias}) gives
rise to potential scattering of the form
\begin{equation}
H_0 = J_{p} \sum_{a \sigma}{\psi_{a \sigma}^\dagger(0)} \psi_{a \sigma}(0)
\end{equation}
the second term in Eq.~(\ref{Ias}) gives the spin-Kondo part
\begin{equation}
H_1 = J_{1} \sum_\alpha   S^\alpha \sum_{a \sigma \sigma'}{\psi^\dagger_{a \sigma}}(0) {\sigma^\alpha}_{\sigma \sigma'}\psi_{a \sigma'}(0),
\end{equation}
the third gives orbital-Kondo part
\begin{equation}
H_2 = J_{2} \sum_a   T^a \sum_{\alpha \sigma \beta}{\psi(0)^\dagger }_{\alpha \sigma} {\lambda^a}_{\alpha, \beta}\psi(0)_{\beta \sigma}
\end{equation}
and the last gives spin-orbit Kondo part
\begin{equation}
H_3 = J_{3} \sum_{a,b}   T^a \otimes S^b \sum_{\alpha \sigma\beta} {\psi(0)^\dagger }_{\alpha \sigma} {\lambda^a}_{\alpha,\beta} {\sigma^b}_{\sigma, \sigma'}\psi(0)_{\beta \sigma'}
\end{equation}

The Kondo-couplings $J_1,J_2,J_3$ depend on the valence $n_f$ and type
of the projector $P$. We first consider the SU(6) symmetric case,
which is realized in the absence of Hund's rule coupling. In this
case, the projector $P$ is irrelevant, since all states at some valence
have equal energy. The local operators are then simply given by
\begin{eqnarray}
&&\sum_{a\sigma\sigma'} f_{a\sigma} \sigma^\alpha_{\sigma \sigma'} f^\dagger_{a\sigma'} = -2S^\alpha
\nonumber\\
&&\sum_{a b\sigma} f_{a\sigma} \lambda^\alpha_{a b} f^\dagger_{b\sigma'} = -T^\alpha
\nonumber\\
&&\sum_{ab\sigma\sigma'} f_{a\sigma} \lambda^\alpha_{a b}
\sigma^\beta_{\sigma \sigma'} f^\dagger_{b\sigma'} = -2 T^\alpha S^\beta,
\nonumber
\end{eqnarray}
and the Kondo couplings become
$J_1 = 2/6\; J_0=J_0/3$, $J_2= J_0/4$ and $J_3= 2/4\; J_0 $, where
$J_0=\frac{V^2}{2U+\varepsilon_f}>0$.
Notice that all Kondo-couplings are positive (minus sign comes from
$\Delta E$ and from the proportionality constant) and hence
antiferromagnetic couplings ensures complete quenching of the spin and
orbital moment. The ground state is thus Fermi liquid.

In the limit of large Hund's coupling, the projector $P_{n+1}$ projects to
the subspace of high-spin states only, which in the case of the three
band model and $n_f=3$, take the following form
\begin{eqnarray}
|1 \rangle &\equiv& |\dn\dn\dn\rangle\label{st1}\\
|2\rangle &\equiv& \frac{1}{\sqrt{3}} (|\dn\dn\up\rangle+|\dn\up\dn\rangle+|\up\dn\dn\rangle)\label{st2}\\
|3\rangle &\equiv& \frac{1}{\sqrt{3}} (|\dn\up\up\rangle+|\up\dn\up\rangle+|\up\up\dn\rangle)\label{st3}\\
|4\rangle &\equiv& |\up\up\up\rangle\label{st4}
\end{eqnarray}
Projection to the ground state multiplet $P_n$ is achieved by
projecting to the following states
\begin{eqnarray}
|5 \rangle& \equiv&  |\dn 0 \dn\rangle\\
|6 \rangle& \equiv&  |0 \dn \dn\rangle\\
|7\rangle& \equiv& |\dn \dn 0\rangle\\
|8\rangle& \equiv& \frac{1}{\sqrt{2}}(|\dn 0 \up\rangle + |\up 0 \dn\rangle)\\
|9\rangle& \equiv& \frac{1}{\sqrt{2}}(|0 \dn \up\rangle + |0 \up \dn\rangle)\\
|10\rangle& \equiv& \frac{1}{\sqrt{2}}(|\dn \up 0\rangle + |\up \dn 0\rangle)\\
|11\rangle& \equiv& |\up 0 \up\rangle\\
|12\rangle& \equiv& |0 \up \up\rangle\\
|13\rangle& \equiv& |\up \up 0\rangle
\end{eqnarray}

An explicit calculation can be used to determine proportionality
constants
\begin{eqnarray}
&&\sum_{a\sigma\sigma'} P_n f_{a\sigma} \sigma^\alpha_{\sigma \sigma'}  P_{n+1} f^\dagger_{a\sigma'} P_n = \frac{2}{3}S^\alpha
\nonumber\\
&&\sum_{a b\sigma} P_n f_{a\sigma} \lambda^\alpha_{a b} P_{n+1} f^\dagger_{b\sigma'}P_n = -\frac{4}{3} T^\alpha
\nonumber\\
&&\sum_{ab\sigma\sigma'} P_nf_{a\sigma} \lambda^\alpha_{a b} \sigma^\beta_{\sigma \sigma'} P_{n+1} f^\dagger_{b\sigma'}P_n = -\frac{4}{3}T^\alpha S^\beta
\nonumber
\end{eqnarray}
We can finally determine the Kondo couplings in the limit of large
Hund's coupling. Their value is $J_1=-2/3 * 1/6 J_0=-J_0/9$, $J_2=4/3*1/4 J_0=J_0/3$
and $J_3 = 4/3*1/4 J_0=J_0/3$. Here $J_0=\frac{V^2}{2U-2J_H+\varepsilon_f}>0$.

The crucial result of this calculation is that the spin-spin Kondo
coupling $J_1$ changes sign when Hund's coupling is strong. This
comes from the fact that the spin operator in the projected subspace
$\sum_{a\sigma\sigma'} P_n f_{a\sigma} \sigma^\alpha_{\sigma \sigma'} P_{n+1} f^\dagger_{a\sigma'} P_n= \frac{2}{3}S^\alpha$
has very different expansion in terms of electron field operator than
in non-projected case
$\sum_{a\sigma\sigma'} f_{a\sigma} \sigma^\alpha_{\sigma \sigma'} f^\dagger_{a\sigma'} = -2S^\alpha$.
The origin of this sign change is in the orbital blocking mechanism,
which ensures that the intermediate state at $n_f=3$ is a high
spin-state (in this case $S=1$) but is orbitally a singlet state,
such as states~\ref{st1}-\ref{st4}.
Orbital blocking is a restriction in the Hilbert space imposed by the
large Hund's rule coupling. It modifies the Kondo couplings away from
their SU(N) symmetric values ($J_1=J_0/3$, $J_2=J_0/4$, $J_3=J_0/2$).
This blocking results in different Kondo
couplings in different valences. For the half-filled shell (relevant
for Mn$^{2+}$) it results in $J_2=0$, $J_3=0$, and a strong reduction
of the value of $J_1$, first recognized by
Schrieffer~\cite{Schrieffer}. For the valence of one unit of charge
away from half-filling (relevant for Fe$^{2+}$ and Ru$^{4+}$), orbital
blocking results in the sign reversal of $J_1$.

\subsection{Results for the three band Hubbard Model}

\begin{figure}[!ht]
\centering{
\includegraphics[width=0.99\linewidth]{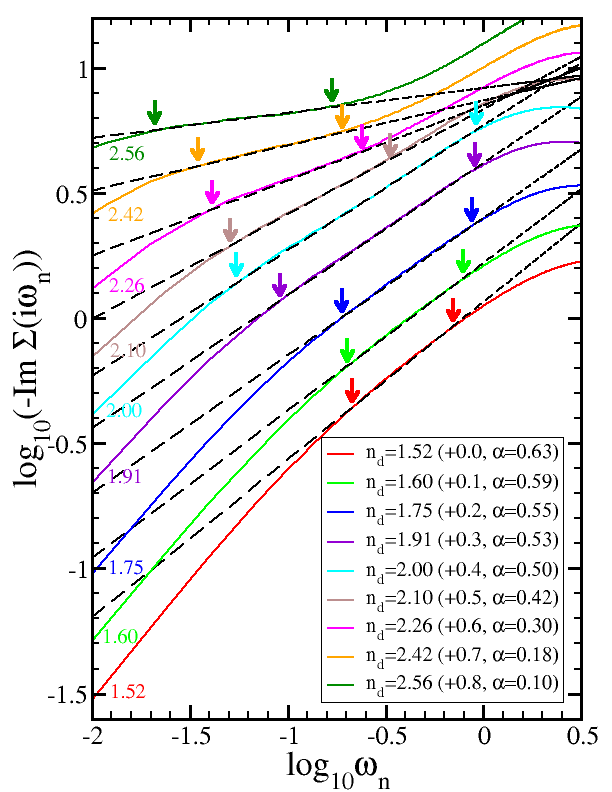}
}
\caption{
(color online) \textbf{Quasiparticle self energy for $J_H$=2.0.}
The imaginary part of the quasiparticle self energy in log$_{10}$-log$_{10}$ scale
as a function of electron occupation $n_d$ for $U$=6.0 and $J_H$=2.0.
Note the data are shifted along the $y$-axis for better illustration.
The linear dispersion of the self energy in the plot indicates the powerlaw behavior exists
in the intermediate frequency region as indicated by the arrows.
}
\label{self1}
\end{figure}

\begin{figure}[t]
\centering{
\includegraphics[width=0.99\linewidth]{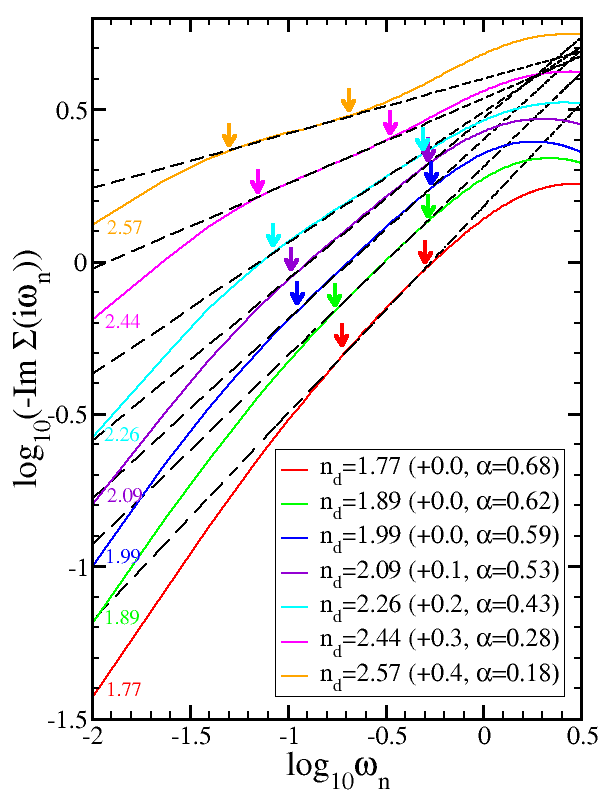}
}
\caption{
(color online) \textbf{Quasiparticle self energy for $J_H$=1.0.}
The imaginary part of the quasiparticle self energy in log$_{10}$-log$_{10}$ scale
as a function of electron occupation $n_d$ for $U$=6.0 and $J_H$=1.0.
Note the data are shifted along the $y$-axis for better illustration.
The linear dispersion of the self energy in the plot indicates the powerlaw behavior exists
in the intermediate frequency region as indicated by the arrows.
}
\label{self2}
\end{figure}

Using our numerical methods of quantum
Monte Carlo, we can not obtain high precision real axis self-energy,
however, we can infer its analytic properties from imaginary axis
analogs. The fractional exponent in scattering rate on the real axis
($Im\Sigma(\omega)\propto \omega^\alpha$) leads to the same powerlaw
on imaginary axis for the imaginary part ($Im\Sigma(i\omega_n)\propto
-\omega_n^\alpha$). The real part, on the other hand, shows the
powerlaw only when scattering rate is very asymmetric around zero
frequency. For example, $\Sigma''(\omega>0)=A|x|^\alpha$ and
$\Sigma''(\omega<0)=B|x|^\alpha$, the real part on imaginary axis is
$Re\Sigma(i\omega_n)\propto (A-B) \int_0^\Lambda x |x|^\alpha
/(x^2+\omega_n^2) dx$ (where $\Lambda$ is the upper cutoff for the
powerlaw), and does not show powerlaw in the symmetric $A=B$ case. Our
calculation shows that the real part does not show very clear powerlaw
on imaginary axis, hence we infer that the scattering rate is quite
symmetric at low frequency on real axis.

Figure \ref{self1} and \ref{self2} show the imaginary part of
the quasiparticle self energy in log$_{10}$-log$_{10}$ scale
as a function of electron occupation $n_d$ for $U$=6.0 and $J_H$=2.0 and 1.0, respectively.
The linear dispersion of the self energy in the plots indicates the powerlaw behavior exists
in the intermediate frequency region as indicated by the arrows.
For both values of $J_H$, the powerlaw exponent $\alpha$ decreases monotonically with
increasing $n_d$ towards half filling, i.e., $n_d$=3.
The upper energy cutoff $\varepsilon_1^*$ drops rapidly for $n_d>$2.0,
suggesting the powerlaw behavior is vanishing quickly when $n_d$ goes away from 2 to half filling.
Compared to $J_H$=2.0, the powerlaw behavior for $J_H$=1.0 is valid in a smaller frequency region and the corresponding
powerlaw exponent is larger, suggesting the important role of Hund's coupling in giving rise to the powerlaw behavior.
Therefore the powerlaw behavior is most visible at
electron occupation one unit of charge away from half-filling, in this case, $n_d=2$

\begin{figure}[t]
\centering{
\includegraphics[width=0.99\linewidth]{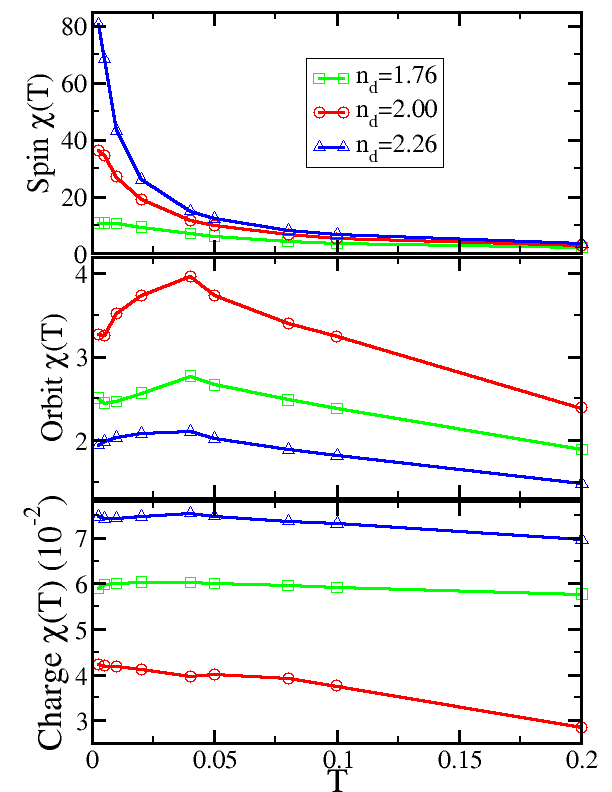}
}
\caption{
(color online) \textbf{Local spin, orbital, and charge susceptibility}
The local spin, orbital and charge susceptibility at zero frequency
as a function of temperature for $n_d$=1.75, 2.00, 2.26 and $U$=6.0, $J_H$=2.0.
The spin susceptibility has large static values and takes the Curie-Weiss form
while the orbital susceptibility is Pauli-like and enhanced at intermediate temperature
and around $n_d$=2.0. Note the charge susceptibility is two orders of magnitude smaller than
the orbital susceptibility thus doesn't play an important role.
}
\label{susc}
\end{figure}

Figure \ref{susc} shows the local spin, orbital and charge susceptibility at zero frequency
as a function of temperature for $n_d$=1.75, 2.00, 2.26 and $U$=6.0, $J_H$=2.0.
The spin susceptibility has large static values and takes the Curie-Weiss form
while the orbital susceptibility is Pauli-like and enhanced at intermediate temperature
and around $n_d$=2.0. Note the charge susceptibility is two orders of magnitude smaller than
the orbital susceptibility thus doesn't play an important role.

\end{document}